\documentclass[aps,showpacs,a4paper,floatfix,twocolumn,prc,amsmath,amssymb]{revtex4-1}
\usepackage{graphicx}
\usepackage{epsfig,hyperref}
\usepackage{color} 
\usepackage{morefloats}
\usepackage{rotating}
\usepackage{relsize}
\usepackage{url}

\usepackage{booktabs}
\usepackage{soul}
\usepackage{array}

\begin{document}

\title{Heavy baryons  in hot stellar matter with light nuclei and hypernuclei}

\author{Tiago Cust{\'o}dio}
\author{Helena Pais}
\author{Constan\c ca Provid{\^e}ncia}

\affiliation{CFisUC, Department of Physics, University of Coimbra,
	3004-516 Coimbra, Portugal.}

\begin{abstract}
	The production of light  nuclei and hypernuclei together with heavy baryons, both hyperons and $\Delta$-baryons, in low density matter as found in stellar environments such as supernova or binary mergers is studied within  relativistic mean-field models. 
	Five light nuclei were considered together	with three light hypernuclei.  The presence of both hyperons and $\Delta$-baryons shift the dissolution of clusters to larger densities and increase the abundance of clusters. This effect is larger the smaller the charge fraction and the higher the temperature.
	The couplings of the $\Delta$-baryons were chosen imposing that the nucleon effective mass remains finite inside neutron stars.
	
\end{abstract}

%-------------------------------------------------------------------------
\maketitle

\section{Introduction}

Neutron stars (NS) are born in highly energetic events called core-collapse supernovae (CCS). Right after the core-collapse, the proto-neutron star reaches high temperatures of the order of tens of MeV. However, in a matter of a few seconds, neutrinos and photons diffuse out of the star, and the star cools down to less than 1 MeV, reaching its ground-state configuration in chemical equilibrium (also known as $\beta$-equilibrium) \cite{Glend}.
The star will remain in equilibrium unless it is perturbed by some external phenomena, such as a collision with another NS.
In these type of events, both CCS and neutron star mergers (NSM), $\beta$-equilibrium is not necessarily achieved, and temperatures as high as 50 to 100 MeV may be attained \cite{Oertel2017}. At such high temperatures, exotic degrees of freedom such as hyperons and $\Delta-$isobars may appear at much lower densities, as compared to the NS case. In fact, a finite temperature allows for the presence of excited states of the nucleons, which can then be converted into heavier baryons at lower densities. Therefore, to describe such events, it is necessary to consider a wide range of charge fractions, temperatures and densities. 

In the NS inner crust, heavy neutron-rich clusters (pasta phases) \cite{Ravenhall1983,Watanabe2005,pais2012PRL} should form, immersed in a gas of neutrons and electrons \cite{chamel2008physics,Vidana2018}. Light clusters, such as $^2$H, $^3$H, $^3$He, $^4$He, $^6$He, are also expected to be present for temperatures above 1 MeV \cite{Pais16}. As the density increases even further, these heavy clusters will eventually melt at densities of $\sim 0.5 n_0$. This sets the transition to the core of the star. In this region, the composition of the star corresponds to uniform nuclear matter made of neutrons, protons, electrons and muons \cite{Vidana2018}. In the inner core of the star (densities of the order of $\sim 2 n_0$), exotic degrees of freedom such as hyperons and delta isobars, or even deconfined quark matter, may appear \cite{Vidana2018}.

Hyperons, together with the nucleons, form the the spin$-1/2$ baryonic octet. $\Delta-$isobars are spin$-3/2$ baryons formed by $u$ and $d$ quarks, that usually decay via the strong force into a nucleon and a pion. These exotic degrees of freedom will appear at high densities, reducing the pressure of the system, when the increasing chemical potentials of the nucleons approach the effective mass of hyperons and $\Delta$s, so that the nucleons start to be converted into these new degrees of freedom  \cite{Glend,Drago2014,KOLOMEITSEV2017,Ribes_2019}.

Besides reducing the Fermi pressure, the introduction of hyperons decreases the free energy of matter \cite{Marques2017,Fortin2017}. These authors also showed that, at low densities, hyperons can compete with light clusters, implying that the minimization of the free energy should also allow for the appearance of hyperons at these densities. In Ref.~\cite{Menezes2017}, the possible appearance of hyperons in the density region of the non-homogeneous matter that forms the inner crust of a NS was analyzed. Temperatures below the melting temperature of the heavy clusters that form this region were considered, i.e $T\lesssim 15$ MeV. It was found that only very small amounts of hyperons, like $\Lambda$ fractions below $10^{-5}$, were present in the background gas. The low-density EoS of stellar matter including light clusters and heavy baryons was also studied in Ref.~\cite{Sedrakian2020}. In addition to hyperons, the author also considered $\Delta-$baryons, pions, and the presence of a representative heavy cluster. It was shown that, depending on temperature and density, the composition of matter may shift from a greater abundance of light clusters to a heavy-baryon predominance.

In a recent work \cite{Custodio2021}, the calculation of the abundance of purely nucleonic light clusters ($^2$H, $^3$H, $^3$He, $^4$He and $^6$He) and hyperclusters ($^3_{\Lambda}$H, $^4_{\Lambda}$H, $^4_{\Lambda}$He) as well as hyperons was performed in the framework of relativistic mean-field models for finite temperature and fixed proton fraction. In the present work, we intend to include $\Delta-$isobars and use two relativistic mean-field models, FSU2H \cite{FSU2H} and DD2 \cite{Typel2009}.
The introduction of clusters is going to follow the approach first presented in Ref.~\cite{Pais2018}, where the effect of the medium on the binding energy of the clusters is considered through the introduction of a binding energy shift, together with a universal coupling of the scalar $\sigma-$meson to the different clusters, that was chosen so that the equilibrium constants of the NIMROD experiment \cite{Qin2011} were reproduced. In Refs.~\cite{Pais2020prl,Pais2020}, this theoretical approach \cite{Pais2018} was applied to the description of the INDRA data \cite{indra}, where an experimental analysis of data was also done including in-medium effects. It was verified that, due to the inclusion of the in-medium effect in the experimental analysis, the equilibrium constants were reproduced with a larger $\sigma-$meson coupling. The calibration of the scalar meson to the clusters coupling was later performed for other models in Ref.~\cite{Custodio2020}.

The structure of this paper is as follows: in the next Section, we briefly describe the formalism used, in Section \ref{sec3}, the results are discussed, and finally, in Section \ref{sec4}, some conclusions are drawn.

\section{Relativistic mean field description of hadronic matter \label{sec2}}

We briefly present the formalism used to describe warm matter which,  besides nucleons, also includes  light clusters, light hyperclusters, both considered as point-like particles,  and heavy baryons, both hyperons and the $\Delta$-baryons. We, therefore,  generalize the study performed in \cite{Custodio2021}, in order to include the $\Delta$ baryons.

The description of the hadronic matter will be carried out within a relativistic mean-field (RMF) approach, see for instance \cite{Glend}.
The interaction is described by the exchange of virtual mesons, in particular,  the following  mesons will be introduced: the usual  isoscalar-scalar $\sigma$ meson, isoscalar-vector $\omega^{\mu}$ and  isovector-vector
$\vec{\rho^{\mu}}$,  together with the isoscalar-vector $\phi^{\mu}$ meson field  with hidden strangeness responsible for an extra
repulsion between hyperons.

We consider both a  model with density-dependent couplings (DD2) and a model with nonlinear meson terms (FSU2H), two different approaches of introducing  the density dependence on the symmetric nuclear matter EoS and on the symmetry energy. Both models satisfy constraints from observations (they are able to reach 2 $M_\odot$ stars), experimental data, and theoretical  calculations.

The light clusters considered in the present study are the usual  purely nucleonic
light nuclei ($^{2}$H, $^3$H, $^3$He, $^4$He) together with the neutron rich nucleus $^6$He. Concerning light hypernuclei nuclei, we introduce the three loosely bound
hypernuclei: the $^{3}_{\Lambda}\text{H}$ hypertriton \cite{Adam2019}, the
$^{4}_{\Lambda}\text{H}$ hyperhydrogen 4 \cite{Esser2014} and the hyperhelium4
$^{4}_{\Lambda}\text{He}$ \cite{Yamamoto2015}.  The spin and isospin of these clusters have been summarized in \cite{Custodio2021}. Only clusters with a charge $Z\le2$ are introduced, because we will essentially consider temperatures above 15 MeV in order to have noticeable fractions of heavy baryons at low densities. Under these conditions, it was shown in \cite{Pais2019}
that heavy clusters have already dissolved and  the light clusters that remain have a charge $Z\le2$.

The approximation introduced concerning the description of light clusters as point-like particles  sets some limitations, in particular,  the density of clusters should be low.   Above a given density, defined by the interaction and the temperature,   clusters melt and  matter is only constituted by nucleons and heavy baryons. In our approach, the $\omega$-meson is the main responsible for this process, although other descriptions are possible such as the introduction of an excluded volume  \cite{LS,Shen1998,Hempel2010,Raduta2010,Furusawa2017}.

The Lagrangian density for this system reads \cite{Glend,Typel2009,Pais2018,Custodio2021}
\begin{equation}\label{Lagrangian_nlm_inicial}
	\mathcal{L}=\!\sum_{\substack{b=baryonic \\ octet, {\Delta}}} \!\!\!\!\!\!\mathcal{L}_b + \!\!\sum_{\substack{i=light \\ nuclei,\\hypernuclei}} \!\!\!\mathcal{L}_{i} 
	+ \!\!\sum_{\substack{m=\sigma,\omega,\phi,\rho }}\!\!\!\!\mathcal{L}_m + \mathcal{L}_{nl}.
\end{equation}
The sum over $b$ extends over the spin-1/2 baryonic octet and the  spin-3/2 $\Delta$ quadruplet. The second term is the sum over light nuclei and light hypernuclei, and the last two terms refer to the mesonic terms, where $\mathcal{L}_{nl}$ includes all the nonlinear mesonic terms and is  only present in FSU2H.

The baryonic term in Eq. (\ref{Lagrangian_nlm_inicial}) reads
\begin{eqnarray}
	\mathcal{L}_b= \bar{\Psi}_b & \left[ i \gamma_{\mu} \partial^{\mu} - m_b + g_{\sigma b} \sigma - g_{\omega b} \gamma_{\mu} \omega^{\mu} \right. \\ & \left. - g_{\rho b} \gamma_{\mu} \vec{I}_b\cdot \vec{\rho^{\mu}} - g_{\phi b} \gamma_{\mu} \phi^{\mu} \right]  \Psi_b \, , \nonumber
\end{eqnarray}
where  $\Psi_b$ is the baryon field, $ \vec{I}_b $ the isospin operator and the  parameters $ g_{mb} $ 
are the couplings parameters  of the baryons to the mesons. 
The  other parameters of the model are the nucleonic vacuum mass  $m=m_n=m_p= $ 939 MeV, the  hyperon masses, $ m_{\Lambda}=1115.683 $ MeV, $ m_{\Sigma^-}=1197 $  MeV, $
m_{\Sigma^0}=1193 $ MeV, $ m_{\Sigma^+}=1189 $ MeV, $ m_{\Xi^-}=1321 $
MeV, and $ m_{\Xi^0}=1315 $ MeV, and the $\Delta$ masses taken to be equal to 1232 MeV.

In Table~\ref{tab1}, we summarize the symmetric nuclear matter properties of the two models considered.
\begin{table}[htb]
\caption{\label{tab1}
The symmetric nuclear matter properties at saturation density for DD2 and FSU2H: the nuclear saturation density $\rho_0$, the binding energy per particle $B/A$, the incompressibility $K$, the symmetry energy $E_{sym}$, the slope of the symmetry energy $L$, and the nucleon effective mass $M^{*}$. All quantities are in MeV, except for $\rho_0$ that is given in fm$^{-3}$, and the effective nucleon mass is normalized to the nucleon mass.}
\begin{ruledtabular}
\vspace{0.5cm}
\begin{tabular}{ccccccc}
 Model    &  $\rho_0$ & $B/A$ &  $K$ & $ E_{sym}$ & $L$ & $M^{*}/M$\\
\hline
FSU2H & 0.15  & 16.28  & 238  &  30.5  & 45 & 0.59  \\
DD2   & 0.149   & 16.02   & 243  &  31.7 & 58 & 0.56 \\
\end{tabular}
\end{ruledtabular}
\end{table}

For the DD2 model  \cite{Typel2009} with density-dependent coupling parameters,
the couplings $ g_{mN} $ of the nucleons ($ N=n,p $) to the $ \sigma$, $\omega$ and $\rho$ mesons are given by 
\begin{equation}
  g_{mN}(n_B) = g_{mN}(n_0) h_M(x)~,\quad x = n_B/n_0~,
\end{equation}
with  $n_B$ the baryonic density, and $n_0$ the saturation density. The isoscalar couplings depend on the function $h_M$ given by ~\cite{Typel2009},
\begin{equation}
h_M(x) = a_M \frac{1 + b_M ( x + d_M)^2}{1 + c_M (x + d_M)^2}
\end{equation}
while for the isovector coupling, $h_M$ has the form
\begin{equation}
h_M(x) = \exp[-a_M (x-1)] ~,
\end{equation}
with the parameters $a_M, b_M, c_M,$ and $d_M$ defined in
Ref.~\cite{Typel2009}. Other parameters of the model such as the meson masses are also given in ~\cite{Typel2009}. For the other  model, FSU2H, the couplings are defined in  \cite{FSU2H}.

The mesonic Lagrangian densities are given by
\begin{eqnarray}
	\mathcal{L}_m &=&
 \frac{1}{2} (\partial_\mu \sigma \partial^\mu \sigma - m_\sigma^2 \sigma^2)\\
&&- \frac{1}{4}
W_{\mu\nu} W^{\mu\nu} 
- \frac{1}{4}
P_{\mu\nu} P^{\mu\nu} 
- \frac{1}{4}
\vec{R}_{\mu\nu} \vec{R}^{\mu\nu} \nonumber \\ && 
+ \frac{1}{2} m^2_\omega \omega_\mu \omega^\mu 
+ \frac{1}{2} m^2_\phi \phi_\mu \phi^\mu 
+ \frac{1}{2} m^2_\rho \vec{\rho}_\mu \cdot \vec{\rho}^\mu ~,
\end{eqnarray}
and
	\begin{eqnarray} \label{Lagrangian_nl}
		\mathcal{L}_{nl}  &=&  - \frac{\kappa}{3!}g_{\sigma N}^3\sigma^3 - \frac{\lambda}{4!}g_{\sigma N}^4\sigma^4  +  \frac{\zeta}{4!}g_{\omega N}^4\omega_{0}^4 \\[8pt]
		&+& \Lambda_{\omega} g_{\rho N}^2  g_{\omega N}^2 \rho_{03}^2 \omega_{0}^2  \nonumber
	\end{eqnarray}
The meson masses and couplings $\kappa$, $\lambda$, $\zeta$, $\Lambda_{\omega}$ are defined for each parametrization, see \cite{Typel2009} for DD2, \cite{FSU2H} for FSU2H.

\subsection{Couplings of mesons to the heavy baryons}\label{secIA}

The couplings of the hyperons  ($ \Lambda,\Sigma^-,\Sigma^0,\Sigma^+,\Xi^-,\Xi^0 $) to the mesons $ g_{mb} $ are defined in terms of the nucleon couplings as $ g_{mb}=x_{mb}\, g_{mN} $, with $ m=\sigma,\,\omega,\,\rho $  and  $ g_{\phi b}=x_{\phi b}\, g_{\omega N} $. In Tables~\ref{tab2} and \ref{tab3}, we give the values considered in the present work.
For the vector mesons, $\omega$ and $\phi$, the ratios $x_{mb}$ are defined according to the SU(6) quark model. For the $\rho$-meson we take $x_{\rho b}=1$, 
and consider that the magnitude of the isospin projection defines the strength of the coupling.
The couplings of the $\Lambda$-hyperon and the $\Xi$-hyperon to the $\sigma$-meson are taken from references \cite{Fortin2017} and \cite{Fortin2020}, and were calibrated   by fitting the experimental binding energy of  hypernuclei.
For the $\Sigma$-hyperon,  we have considered  that the $\Sigma$ potential in symmetric nuclear matter  is repulsive and equal to $ U_{\Sigma}^{(N)}(n_0)=30 $ MeV, a value commonly used and within the range of values indicated by experiments \cite{Gal_sigma_potential}.

\begin{table}[htb]
\caption{\label{tab2}
Coupling constants of the vector mesons ($m=\omega,\phi,\rho$) to the different hyperons, normalized
to the respective meson nucleon coupling, i.e. $x_{mb}=g_{mb}/g_{mN}$,
except for the $\phi-$meson where the $g_{\omega N}$ is used for normalisation.}
\begin{ruledtabular}
\begin{tabular}{cccc}
	$ b $	& $ x_{\omega b} $   & $x_{\phi b} $            & $ x_{\rho b} $ \\ 
	\hline
		$ \Lambda $ &  2/3 & $ -\sqrt{2}/3 $  & 1   \\
	    $ \Sigma $ &  2/3 & $-\sqrt{2}/3 $  & 1   \\ 
		$ \Xi $ &  1/3 &$ -2\sqrt{2}/3 $  & 1  \\ 
	\end{tabular}
	\end{ruledtabular}
\end{table}

\begin{table}[htb]
\caption{\label{tab3}
Coupling constants of the $\sigma$ meson to the different hyperons, normalized to the $\sigma$ meson nucleon coupling, i.e. $x_{\sigma b}=g_{\sigma b}/g_{\sigma N}$, for the DD2 and FSU2H models.}
\begin{ruledtabular}	
	\begin{tabular}{ccc}
		$ x_{\sigma b} $     & DD2    & FSU2H  \\ \hline \vspace{-0.3cm}  \\
		$ x_{\sigma \Lambda} $ & 0.621 &  0.620 \\
		$ x_{\sigma \Sigma} $  & 0.474  & 0.452 \\
		$ x_{\sigma \Xi} $      & 0.320  & 0.310  \\  
	\end{tabular}
\end{ruledtabular}	
\end{table}

Similarly to what we have done for the hyperons, we can write the couplings of the $ \Delta $ particles to the mesons, $ g_{m \Delta} $, in terms of the nucleon couplings as:
\begin{eqnarray}
	g_{m \Delta}= x_{m \Delta} ~g_{m N} ~,~ m=\sigma,\omega,\rho ~,
\end{eqnarray}
with $ x_{m \Delta} $ being the ratio between the $\Delta$ and nucleon couplings to the mesons.

Due to limited experimental observations, the couplings of the $\Delta$ particles to the mesons are still poorly constrained. Some phenomenological analyses from pion-nucleus scattering \cite{Nakamura_pion_nucleus_scattering}, electron scattering on nuclei \cite{Koch_electron_scattering_on_nuclei} and electromagnetic excitations of the $\Delta$ particles \cite{WEHRBERGER1989}  have set the following constraints on the values of the coupling constants, as summarized in Ref. \cite{Ribes_2019}: (i) the $\Delta$ potential in nuclear matter could be slightly more attractive than the nucleon potential implying that the ratio $ x_{\sigma \Delta} $ should be above 1;
(ii)  $ x_{\sigma \Delta} $ is larger than $ x_{\omega \Delta} $: 
	\begin{eqnarray}
		0\lesssim x_{\sigma \Delta} - x_{\omega \Delta} \lesssim 0.2 ~; \label{x_sigma_minus_x_omega}
	\end{eqnarray}
(iii) there are no experimental constraints on $ x_{\rho \Delta} $.
We will take into account the  uncertainties associated to these couplings, allowing the couplings to vary within a large interval of values, as done by other authors \cite{Drago2014,KOLOMEITSEV2017,Ribes_2019,Raduta2021}. In the present work,  we will adopt the following intervals  
\begin{eqnarray}
	0.9\leq &x_{\sigma \Delta}&\leq 1.2 \label{x_sigma_delta_interval}\\ [5pt]
	0.9\leq &x_{\omega \Delta}&\leq 1.2 \label{x_omega_delta_interval}
\label{delta_couplings_interval_values} ~.
\end{eqnarray}
The lower limit is set to 0.9 because the nucleon effective mass goes to zero at quite low densities for smaller values. Stronger constraints will be defined by imposing that the effective mass cannot go to zero at densities below the central  density of the maximum mass star. For the coupling to the $\rho$-meson we will consider $x_{\rho \Delta}= 0.8, \, 1,$ and 2.

\subsection{Chemical equilibrium}

In order to impose chemical equilibrium, the chemical potentials of baryons and light clusters and hyperclusters are needed.
The chemical potential of baryon $b$ is given by
	\begin{equation}\label{energy_eigenvalues_ddm_bar_octet}
	\mu_b= g_{\omega b} \omega_{0} + g_{\rho b} I_{3b}\rho_{03} + g_{\phi b} \phi_{0} + \Sigma_{0}^R + \sqrt{k_{Fb}^2+ m^{*2}_{b} }~,
	\end{equation}
	where $k_{Fb}$ is the Fermi momentum of baryon $b$, and $\Sigma_{0}^R$ is the rearrangement term, only present in the DD2 model, defined as 
	\begin{eqnarray}\label{rearr_term_no_clusters}
		\Sigma_{0}^R&=& \sum_{c}\Big(\frac{\partial g_{\omega c}}{\partial \varrho}\omega_0 \rho_c + I_{3c}\frac{\partial g_{\rho c}}{\partial \varrho}\rho_{03} \rho_c  + \frac{\partial g_{\phi c}}{\partial \varrho}\phi_0 \rho_c  \nonumber\\ &&-  \frac{\partial g_{\sigma c}}{\partial \varrho}\sigma \rho_c^s \Big) ~,
	\end{eqnarray}
where the sum is over the baryons, $c=b$, for $T=0$ MeV, and over the baryons and clusters, $c=b,i$, for finite temperatures.	
The effective chemical potentials $ \mu_b^* $ that enter the Fermi distributions are given by 
	\begin{equation}\label{mub*}
		\mu_b^*= \mu_b - g_{\omega b} \omega_{0} - g_{\rho b} I_{3b}\rho_{03} - g_{\phi b} \phi_{0} -\Sigma_{0}^R ~.
	\end{equation}

In the following, we consider matter in equilibrium with a fixed charge fraction $Y_Q$. All particle chemical potentials can be written in terms of the chemical potentials corresponding to  two conserved charges, baryonic charge and electrical charge. We consider that the strangeness chemical potential is zero, because the  weak force does not conserve strangeness. Therefore the chemical potential of each particle $c$ is  a linear combination of the baryon and electric charge chemical potentials:
	\begin{equation}\label{Chemical_potentials_general_formula}
		\mu_c= b_c \mu_n - q_c \mu_e
	\end{equation}
	where $ b_c $ is the baryon number of particle $ c $; $ q_c $ is the electrical charge (in units of $ +e $); and $ \mu_n $, $\mu_e$ the baryon and electrical charge chemical potentials, respectively.
	Since $\mu_e\!=\!\mu_n\!-\!\mu_p$, the hyperon chemical potentials can be written in terms of the nucleon chemical potentials: 
	\begin{eqnarray}
		\mu_{\Lambda}&=&\mu_n \label{Lambda_chemical_potential}\\ [8pt] 
		\mu_{\Sigma^-}&=&2\mu_n - \mu_p, \hspace{0.3cm} \mu_{\Sigma^0}=\mu_n, \hspace{0.3cm} \mu_{\Sigma^+}=\mu_p \label{Sigmas_chemical_potential}\\  [8pt] 
		\mu_{\Xi^-}&=&2\mu_n - \mu_p, \hspace{0.3cm}  \mu_{\Xi^0}=\mu_n~. \label{Cascades_chemical_potential}
	\end{eqnarray}
	Similarly, for the $\Delta$ isobars we have:
	\begin{eqnarray}
		\mu_{\Delta^-}&=&2\mu_n - \mu_p, \hspace{0.3cm} \mu_{\Delta^0}=\mu_n, \nonumber\\ \mu_{\Delta^+}&=&\mu_p , \hspace{0.3cm} \mu_{\Delta^{++}}=2\mu_p - \mu_n~. \label{Deltas_chemical_potential}
	\end{eqnarray}
	
	In equilibrium, the cluster or hypercluster $ i $ chemical potential  must satisfy
	\begin{equation}\label{muj}
		\mu_{i}=N_i\mu_n + Z_i\mu_p + \Lambda_i \mu_{\Lambda} ~.
	\end{equation}
The corresponding  effective chemical potential $\mu_i^*$ is given by
	\begin{equation}\label{mu*j}
		\mu_i^*= \mu_i - g_{\omega i}\omega_0 - g_{\rho i}I_{3i}\rho_{03} - g_{\phi i}\phi_0 -\Sigma_{0}^R ~.
	\end{equation}
The terms involving $\Lambda_i$ in Eq.(\ref{muj})	and the meson field $\phi_0$ in Eq.(\ref{mu*j}) only contribute to the chemical potential of hypernuclei.
	
The total charge fraction $ Y_Q $  of the system is defined as
	\begin{equation}\label{charge_fraction_with_hyperclusters}
		Y_Q=\sum_{b}q_b\,Y_b  +\sum_{i}\frac{q_i}{A_i}\,Y_i 
	\end{equation} 
	where $q_b$ and $ q_i $ are the electric charge (in units of +$ e $) of baryon $b$ and light cluster or hypercluster $ i $. The mass fraction $ Y_i $ of the cluster or hypercluster  $ i $ is given by 
	\begin{eqnarray}\label{yj}
		Y_i&=&A_i\frac{\rho_i}{n_B} ~.
	\end{eqnarray}

 The introduction of light clusters follows the formalism first presented in \cite{Pais2018}, and the inclusion of hyperons in these clusters, termed hyperclusters, was introduced in a recent work \cite{Custodio2021}, where the details on the calculations can be found.

\section{Results \label{sec3}}
In the present section we present our main results. In the first subsection we constrain the $\Delta$-meson couplings  imposing that the effective mass must be non-zero. In a second subsection, we analyse the effect of the heavy baryons on the cluster abundances.

\subsection{Constraining the $\Delta$ couplings}

In order to introduce the $\Delta$-isobars, $\Delta^{-,0,+,++}$, it is necessary to constraint the $\Delta$ couplings to the mesons. The uncertainty on the $\Delta$ couplings can be accounted for by allowing them to vary within a large interval of values. As explained in Sec.\ref{secIA}, we take them to be  $0.9\leq x_{\sigma \Delta} \leq 1.2;0.9\leq x_{\omega \Delta}\leq 1.2 ; x_{\rho \Delta}=0.8,1, 2$ .

Further constraints are obtained from observations: the EoS must be able to describe 2 $M_\odot$ stars, and the effective mass of nucleons must remain finite inside the star. 
In order to build a complete EoS, it is necessary to match the core EoS to the  crust EoS.
We have considered for the  outer crust the BPS EoS \cite{bps}, and for the inner crust, we take the inner crust EoS obtained within a Thomas-Fermi calculation, including non-spherical heavy clusters  \cite{Grill2014,Pais2015}.  The inner crust EoS for DD2 \cite{DD2inner} and for FSU2H \cite{FSU2Hinner} can be found in the CompOSE database \cite{compose}, an online, free and public repository for EoS. 
For the core EoS,  the full spin-1/2 baryonic octet, the four $\Delta$ isobars, electrons and muons were included in  $\beta$-equilibrium and at $ T=0 $ MeV.

\begin{figure*}%[tp]
	\includegraphics[width=0.9\linewidth]{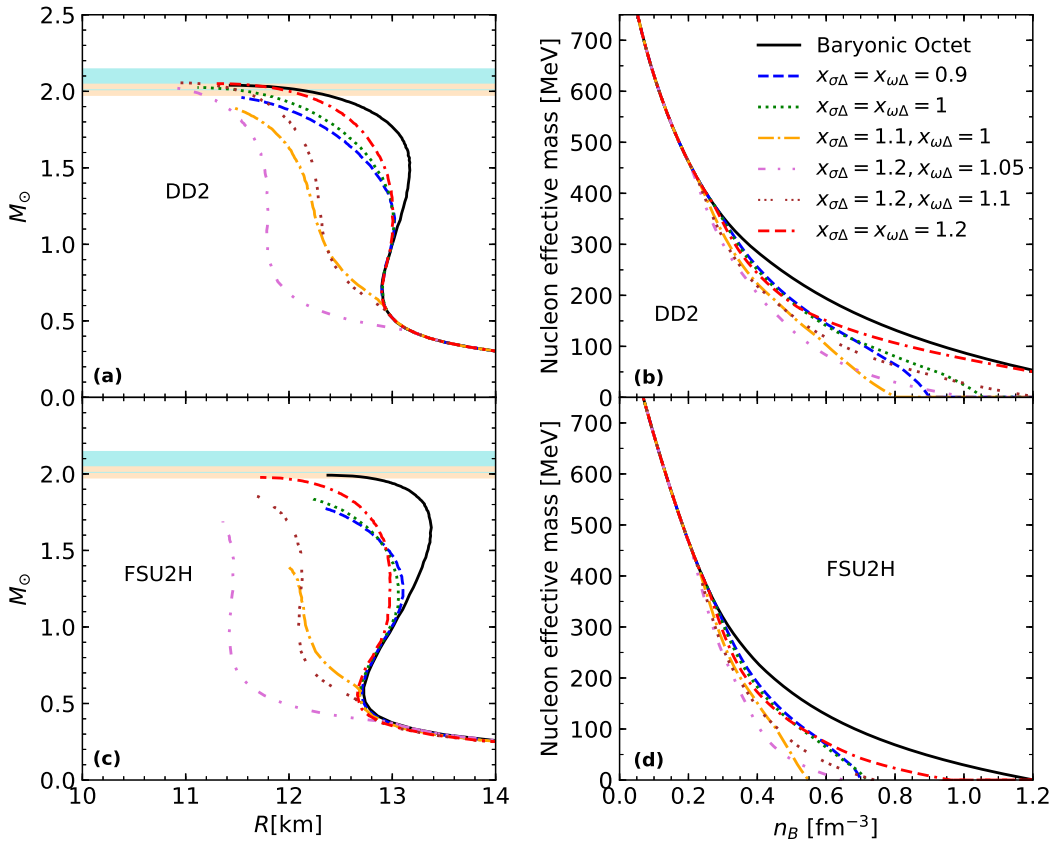}
	\caption{Mass-Radius relations (left) and nucleon effective mass  as a function of the density (right) for several sets of values for the $ x_{\sigma \Delta} $, $ x_{\omega \Delta} $ couplings, fixing $ x_{\rho \Delta}=1 $, and the two RMF models considered, DD2 and FSU2H. The black curves correspond to the EoS without the $\Delta$-baryons. The horizontal bands indicate the mass uncertainties associated to the PSR J0740+6620 \cite{Cromartie2019,Miller_2021} (upper, blue) and PSR J0348+0432 \cite{J0348} (lower, brown) masses.	}
	\label{fig1}
\end{figure*}

In the left panels of Fig.~\ref{fig1} we plot the Mass-Radius relations obtained with the EoSs corresponding to a few representative sets of the $\Delta$ couplings: $ x_{\sigma \Delta}=x_{\omega \Delta}=0.9 $; $ x_{\sigma \Delta}=x_{\omega \Delta}=1 $; $ x_{\sigma \Delta}=1.1,~x_{\omega \Delta}=1 $; $ x_{\sigma \Delta}=1.2,~x_{\omega \Delta}=1.05 $; $ x_{\sigma \Delta}=1.2,~x_{\omega \Delta}=1.1 $; $ x_{\sigma \Delta}=x_{\omega \Delta}=1.2 $. $ x_{\rho \Delta} $ is fixed to 1. For comparison, we also show (in black) the EoS without  $\Delta$-isobars. We also include two horizontal bands corresponding to two of the most massive pulsars ever observed, PSR J0740+6620 \cite{Cromartie2019,Miller_2021},  and PSR J0348+0432 \cite{J0348}, with masses $ M=2.08\pm0.07 M_{\odot} $ and $ M=2.01\pm0.04 M_{\odot} $, respectively.

\begin{figure*}%[tp]
	\includegraphics[width=0.9\linewidth]{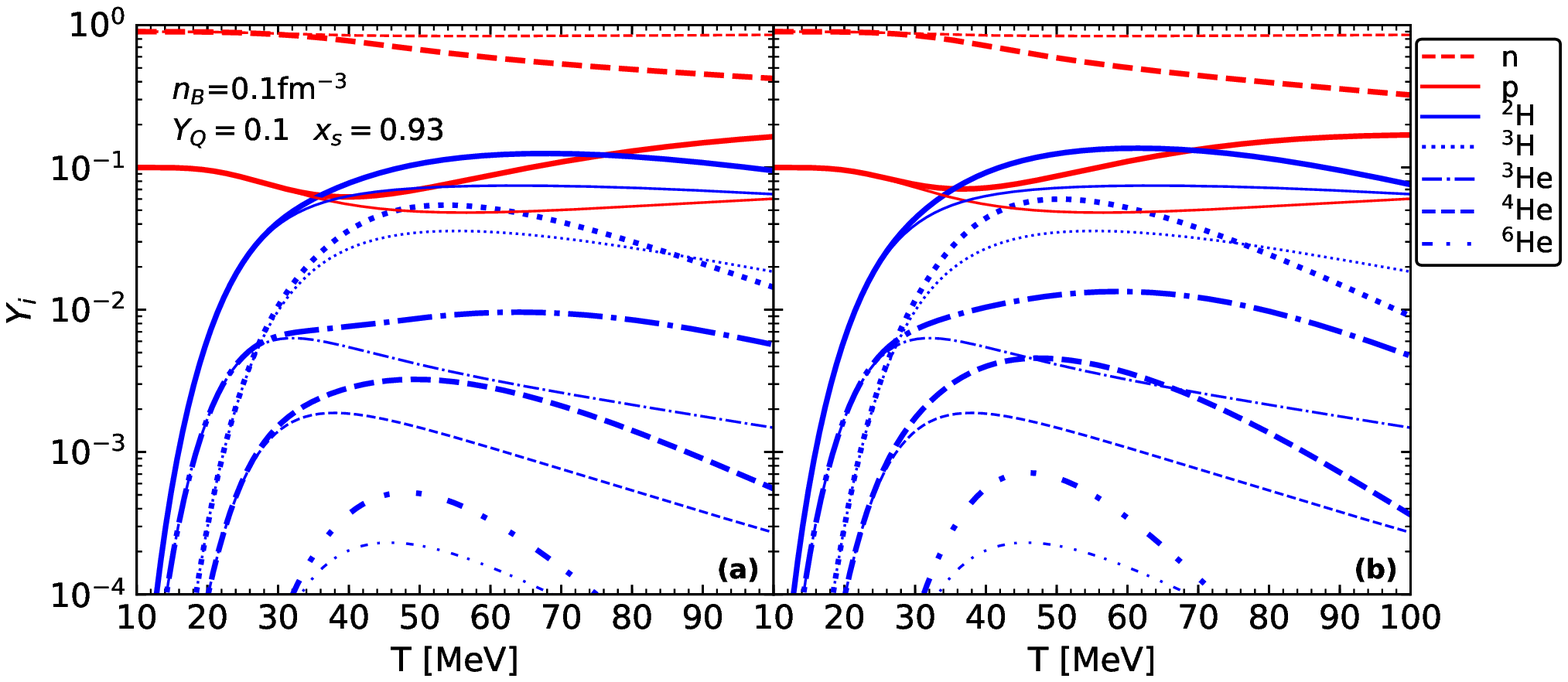}
	\caption{Unbound nucleon (red) and light cluster (blue) fractions for the DD2 model in a calculation with (thick lines) and without (thin lines)  hyperons (a, left), and hyperons and $\Delta$-isobars (b, right), as a function of the temperature for a fixed charge fraction of	$Y_Q=0.1$, and density of $n_B=0.1$ fm$^{-3}$. The scalar cluster-meson coupling is fixed to $x_s=0.93$. The $\Delta$ couplings are fixed to $ x_{\sigma \Delta}= x_{\omega \Delta}=x_{\rho \Delta}=1$.	}
	\label{fig2}
\end{figure*}

For both models, the EoSs which include the $\Delta$ isobars  show a significant decrease in the radius of the intermediate-mass stars ($\sim 1.4 M_{\odot} $), compared to the stars without $\Delta$s. The appearance of the $\Delta$s softens the EoS, and, therefore, the star is further compressed due to gravity and, consequently, has a smaller  radius, as discussed in Ref.\cite{Ribes_2019}. On the other hand, the presence of the $\Delta$s does not seem to significantly affect the maximum masses.

There is also a visible difference in the intermediate-mass radius between the EoSs with different values of the $\Delta$ couplings to mesons, denominated $\Delta x=x_{\sigma \Delta} - x_{\omega \Delta} $.  The EoSs with the same difference in the $\Delta x$  have similar intermediate-mass radii. This is easily understood if we consider the $\Delta$ potential in nuclear matter:
\begin{eqnarray}
	U_{\Delta}^{N}(\rho_0)=-g_{\sigma b} \sigma  + g_{\omega b} \omega_0  + g_{\rho b} I_{3\Delta} \rho_{03}  \label{Potential_Delta} ~.
\end{eqnarray} 
If $ g_{\sigma b} $ increases, the $\Delta$ potential decreases and becomes more attractive, which leads to an  increase in the abundances of the  $ \Delta $s. On the other hand, if $ g_{\omega b} $ increases, the potential becomes less attractive and the $\Delta$s are less favored. Therefore, the larger the $ \Delta x$, the higher the abundances of the $ \Delta $s and the earlier their onset (see Table \ref{tab4}).  An earlier onset  of $\Delta$s  and a higher abundance leads to a more significant softening of the EoS and a larger reduction of its intermediate-mass radius. Looking at Eq.(\ref{Potential_Delta}), it is also reasonable to conclude that different EoSs with the same difference  $ \Delta x$  (e.g. $ x_{\sigma \Delta}= x_{\omega \Delta}=0.9$ and $ x_{\sigma \Delta}= x_{\omega \Delta}=1$) will have similar potentials and, therefore, similar intermediate-mass radii since the increase in $ x_{\sigma \Delta}$ is approximately compensated by a similar increase in $ x_{\omega \Delta}$.

Some of the Mass-Radius relations in the left panels of Fig.~\ref{fig1} do not reach the maximum mass star. The nucleon effective mass corresponding to the EoSs with $\Delta$s decreases much faster than the ones without $\Delta$s, and eventually becomes zero, see  right panels of Fig.~\ref{fig1}. For some parametrizations the effective mass  drops  so fast that it becomes zero before   the maximum mass star is reached. These EoSs are not appropriate to describe NSs   if no phase transition to quark matter is considered, and, therefore, they will be discarded in the present study.

Considering the couplings tested for the DD2  model satisfying Eqs. (\ref{delta_couplings_interval_values}),  the following sets of $\Delta$ couplings are not valid: $ x_{\sigma \Delta}= x_{\omega \Delta}=0.9$; $ x_{\sigma \Delta}=1.1,~ x_{\omega \Delta}=1$; $ x_{\sigma \Delta}=1.2,~ x_{\omega \Delta}=1.05$. For the FSU2H model only the pair  $ x_{\sigma \Delta}= x_{\omega \Delta}=1.2$ corresponds to a valid EoS.   The valid $\Delta$ couplings, i.e., the ones that are able to reach the maximum mass before the nucleon effective mass becomes zero, are shown in Table~\ref{tab4}.

\begin{table*}%[t] 
	\caption{\label{tab4} 
	Maximum mass $ M_{\text{max}} $, correspondent radius $ R(M_{\text{max}}) $, central density $ \rho_c $, $1.4 M_\odot $ radius  $R(1.4M_{\odot}) $, onset density of $\Delta^-$ $ \rho_{\Delta^-} $, onset density of the $\Lambda$ hyperon $\rho_{\Lambda} $, for the DD2 and FSU2H models and considering only the valid $\Delta$ couplings.}
	\centering
	\resizebox{\textwidth}{!}{	
		\begin{tabular}{ccccccc}
			\toprule[0.04cm]
			\toprule \vspace{-0.3cm} \\
			\textbf{DD2}           & $M_{\text{max}}(M_{\odot})$         & $R(M_{\text{max}})$(km)      & $ \rho_c (\text{fm}^{-3}) $      & $ R(1.4M_{\odot}) $(km)      &  $ \rho_{\Delta^-} (\text{fm}^{-3})$ & $ \rho_{\Lambda} (\text{fm}^{-3})$ \\
			\midrule[0.02cm]
			Baryonic Octet 	         & 2.04 & 11.45 & 0.99 & 13.91 & -                     & 0.33              \vspace{0.05cm}        \\
			
			$x_{\sigma\Delta} = x_{\omega\Delta} = 1, x_{\rho\Delta} =1$	         & 2.02 & 11.11 & 1.05 & 12.93 & 0.28 & 0.36  \vspace{0.05cm} \\
			
			$x_{\sigma\Delta} =1.2, x_{\omega\Delta} = 1.1, x_{\rho\Delta} =1$	 	   & 2.06 & 10.95 & 1.05 & 12.26 & 0.23                     & 0.39                \vspace{0.05cm}         \\
			
			$x_{\sigma\Delta} =x_{\omega\Delta} = 1.2, x_{\rho\Delta} =1$     & 2.05 & 11.31 & 1.01 & 12.97 & 0.27                     & 0.35               \vspace{0.05cm}        \\
			$x_{\sigma\Delta} =x_{\omega\Delta} = 1.2, x_{\rho\Delta} =2$     & 2.04 & 11.32 & 1.01 & 13.13 & 0.32                     & 0.34            \vspace{0.05cm}           \\
			$x_{\sigma\Delta} =x_{\omega\Delta} = 1.2, x_{\rho\Delta} =0.8$   & 2.05 & 11.31 & 1.00 & 12.92 & 0.26                     & 0.36            \vspace{0.01cm}             \\
			\bottomrule[0.02cm]
			\toprule[0.02cm]
			\textbf{FSU2H}         & $M_{\text{max}}(M_{\odot})$         & $R(M_{\text{max}})$(km)      & $ \rho_c (\text{fm}^{-3}) $      & $ R(1.4M_{\odot}) $(km)      & $ \rho_{\Delta^-} (\text{fm}^{-3})$ & $ \rho_{\Lambda} (\text{fm}^{-3})$ \\ \midrule[0.02cm]
			Baryonic Octet 	         & 1.99 & 12.39 & 0.79 & 13.29 & -                     & 0.33                     \vspace{0.05cm}         \\
			$x_{\sigma\Delta} =x_{\omega\Delta} = 1.2, x_{\rho\Delta} =1$     & 1.98 & 11.73 & 0.91 & 12.97 & 0.26                     & 0.35                     \vspace{0.05cm}         \\
			$x_{\sigma\Delta} =x_{\omega\Delta} = 1.2, x_{\rho\Delta} =2$     & 1.97 & 11.97 & 0.87 & 13.26 & 0.30                     & 0.34                     \vspace{0.05cm}         \\
			$x_{\sigma\Delta} =x_{\omega\Delta} = 1.2, x_{\rho\Delta} =0.8$   & 1.98 & 11.72 & 0.91 & 12.90 & 0.25                     & 0.36     \vspace{0.01cm}         \\
			\bottomrule   
			\bottomrule[0.04cm]     
		\end{tabular}
	}
\end{table*}

\begin{figure}%[tp]
	\includegraphics[width=1\linewidth]{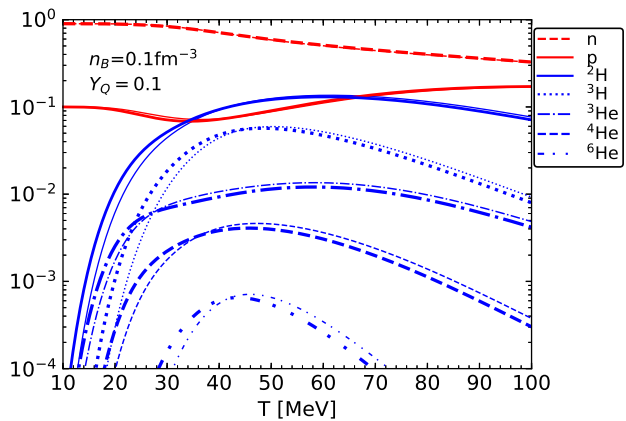}
	\caption{ Unbound nucleon and light cluster fractions as a function of the temperature in a calculation with hyperons and $\Delta$-isobars for the DD2 (thin lines) and FSU2H  (thick lines) models, and a fixed charge fraction of	$Y_Q=0.1$, and density of $n_B=0.1$ fm$^{-3}$. The $\Delta$ couplings are fixed to $ x_{\sigma \Delta}= x_{\omega \Delta}=1.2,~x_{\rho \Delta}=1$.}
	\label{fig3}
\end{figure}

\begin{figure}%[tp]
	\includegraphics[width=1\linewidth]{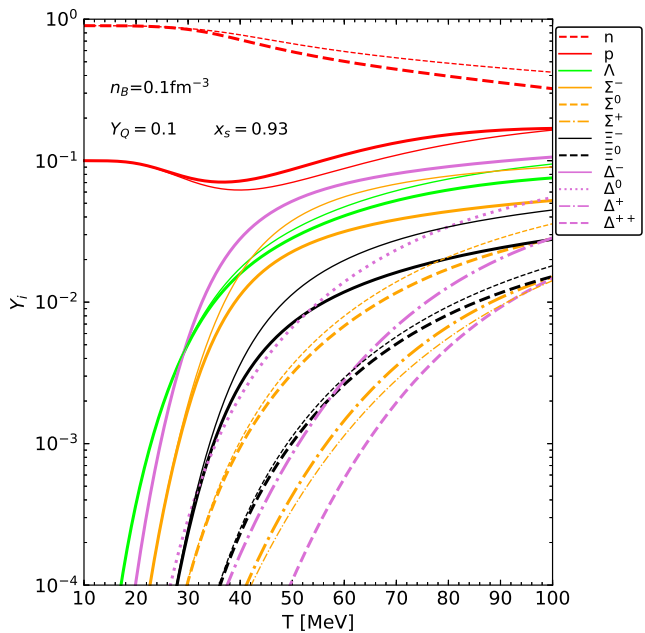}
	\caption{Unbound nucleon and hyperon  fractions as a function of the temperature in a calculation with (thick lines) and without (thin lines) $\Delta$ isobars for the DD2 model, and a fixed charge fraction of $Y_Q=0.1$ and density $n_B=0.1$ fm$^{-3}$. The scalar cluster-meson coupling fraction is set to $x_s=0.93$. The $\Delta$ abundances are also displayed with thick pink lines. Light clusters are present in the calculation but their fractions are not shown. 
	The $\Delta-$couplings are fixed to $ x_{\sigma \Delta}= x_{\omega \Delta}=x_{\rho \Delta}=1$.}
	\label{fig4}
\end{figure}

\subsection{Clusterized matter with $\Delta$-baryons}

In the following we discuss the effect of the presence of $\Delta$-baryons on the properties of clusterized matter for temperatures above 10 MeV, when heavy clusters are not expected anymore. We are mainly going to work with the DD2 model with the $\Delta$ couplings equal to the nucleons,  i.e. $x_{\sigma\Delta} =x_{\omega\Delta}=x_{\rho\Delta} = 1$. However, whenever we compare the DD2 and FSU2H models we have to set $x_{\sigma\Delta} =x_{\omega\Delta} = 1.2$, since these are the only FSU2H valid couplings that remained from the initial six sets of $\Delta$ couplings.

In the left panel of  Fig.~\ref{fig2}, we show the nucleon and purely nucleonic light cluster abundances with (thick lines) and without (thin lines) hyperons as a function of the temperature for a charge fraction of $Y_Q=0.1$ and a density of $n_B=0.1$ fm$^{-3}$.  The right panel shows the same calculation but for a system including $\Delta$-isobars as well, with the correspondent couplings the ones of the nucleons, $ x_{\sigma \Delta}= x_{\omega \Delta}=x_{\rho \Delta}=1$. The inclusion of hyperons increases the fraction of light clusters above $ T=25 $ MeV as discussed in \cite{Custodio2021}.  The presence of both hyperons and $\Delta$s increases even further the abundances of light clusters. This is justified by the reduction of  the  nucleon density in the presence of hyperons and $\Delta$s which leads to smaller binding energy shifts.

\begin{figure*}%[tp]
	\includegraphics[width=0.9\linewidth]{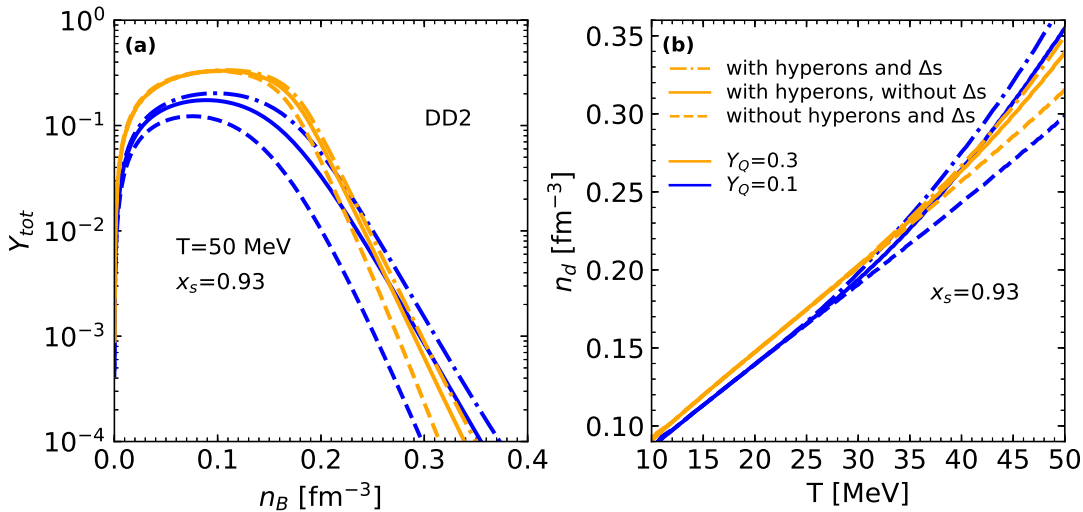}
	\caption{Total mass fraction $Y_{tot}$ of the light clusters  as a function of the density at $T=50$ MeV (left) and the dissolution density of the clusters, $n_d$, as a function of the temperature (right) for a calculation without hyperons and $\Delta$s (dashed), with hyperons and without $\Delta$s (solid), and with hyperons and $\Delta$s (dash-dotted). The charge fraction is fixed to $Y_Q=0.3$ (orange) and 0.1 (blue), and the DD2 model is considered.}
	\label{fig5}
\end{figure*}

In Fig. \ref{fig3}, we show again the abundances of clusters and unbound nucleons in a system with $\Delta$-isobars, 
but for the two models considered in this work: DD2 (thin lines) and FSU2H (thick lines). The scalar cluster-meson coupling is fixed to $x_s= 0.91$ for the FSU2H model, whereas for DD2 is fixed to 0.93 \cite{Custodio2020}.
Although there are some small differences, the overall behavior is very similar. The main difference is the larger abundances of light clusters at low temperatures obtained within FSU2H.

The hyperon abundances are also affected by the inclusion of the $\Delta$ isobars, as clearly seen in Fig.~\ref{fig4}, where  we plot the unbound nucleon and hyperon fractions as a function of the temperature with (thick lines) and without (thin lines) $\Delta$s for a charge fraction of $Y_Q=0.1$ and density $n_B=0.1$ fm$^{-3}$.  The $\Delta$ abundances are also displayed with thick pink lines. The main effect of introducing $\Delta$s is a reduction of the neutrons as well as of the neutral and negatively charged hyperons, whereas the abundances of protons and positively charged $\Sigma^+$ hyperon increase. The most abundant $\Delta$-isobar is clearly the $\Delta^-$ which is negatively charged, so its appearance is compensated by a reduction of the neutral and negatively charged particles and an increase of the positively charged ones. Except for the neutrons, all particles increase their abundances with the temperature. At finite temperature  new channels open and  the interaction, the mass and the charge define the abundances.  It is energetically favorable to convert highly energetic neutrons into other particles.  The more attractive couplings of the $\Delta$s compared with the hyperons  explains why they are more abundant than their equally charged hyperon counterparts.

\begin{figure*}%[tp]
	\includegraphics[width=0.9\linewidth]{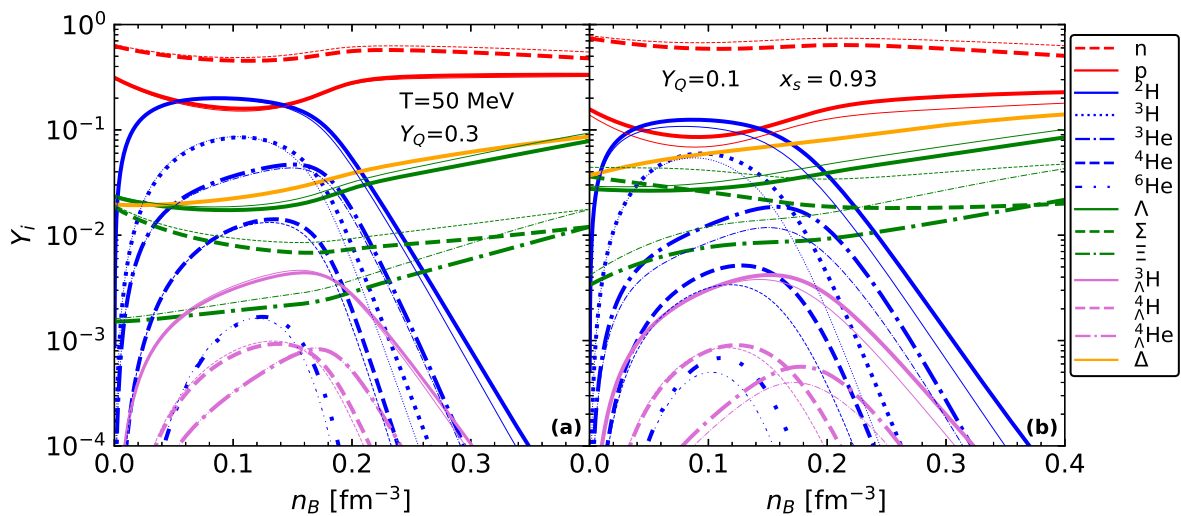}
	\caption{Mass fractions of the unbound nucleons (red), $\Lambda$, $\Sigma$ and $\Xi$ hyperons (green), $\Delta$-isobars (orange), light clusters (blue), and 	light hypernuclei (pink) with (thick) and without (thin) $\Delta$-particles as a function of the density for $T=50$ MeV and  $x_s=0.93$, with  $Y_Q=0.3$ (left) and 0.1 (right). The calculation is performed for the DD2 RMF model. }
	\label{fig6}
\end{figure*}
 
In Fig.~\ref{fig5},  we show the impact on the total mass fractions and dissolution densities of the clusters caused by the inclusion of hyperons only (solid lines) and by the inclusion of both hyperons and  $\Delta$s (dash-dotted lines) for two charge fractions $Y_Q=0.3$ (orange) and 0.1 (blue), and for the DD2 model. The dashed lines were obtained for nucleonic matter, and have been included for comparison. Like in previous Figs, the scalar cluster-meson coupling fraction is set to $x_s=0.93$, and the $\Delta$ couplings are fixed to $ x_{\sigma \Delta}= x_{\omega \Delta}=x_{\rho \Delta}=1$. On the left panel, the behavior of the total cluster mass fraction is plotted for $T=50$~MeV and two charge fractions as a function of density. For $Y_Q=0.3$, the largest cluster fractions are reached for densities below saturation density. Besides, in these range of densities the three different scenarios consider do not differ much. The larger differences occur for $\rho\gtrsim0.2$~fm$^{-3}$, for densities above the maximum in the cluster distribution and close to the dissolution. The presence of the heavy baryons shifts the dissolution density to larger densities. This effect is present considering only hyperons but it is intensified when $\Delta$-baryons are also included. The presence of $\Delta$s reduces the nucleon fraction, and this  is reflected on the medium effects felt by the clusters through a smaller binding energy shift.  For the smaller charge fraction, $Y_Q=0.1$, the difference between the distributions occurs also for densities at the abundance peak , with the largest mass fractions occurring for matter with $\Delta$s and hyperons, followed by matter with hyperons, and the smallest fraction for nucleonic matter. It is clear that the smaller the charge fraction, the larger the difference between the three distributions. The right panel of Fig.~\ref{fig5} summarizes the effect of the heavy baryons on the dissolution density of clusters: the differences occur for temperatures above 25 MeV, with the largest dissolution densities occurring for matter with the smallest charge fraction and containing both hyperons and $\Delta$-baryons. These effects are all understood by realizing that the presence of heavy baryons  reduces the nucleonic background gas, and, therefore, the binding energy shift, preventing clusters to melt.

\begin{figure}%[tp]
	\includegraphics[width=1\linewidth]{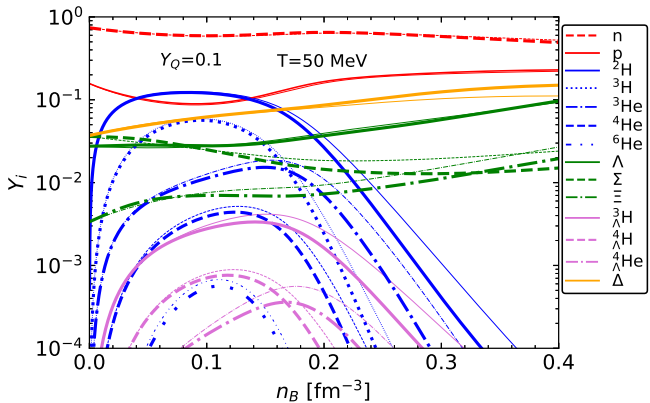}
	\caption{Mass fractions of the unbound nucleons (red), $\Lambda$, $\Sigma$ and $\Xi$ hyperons (green), $\Delta$-isobars (orange), light clusters (blue) and light hypernuclei (pink), with $\Delta$ particles within the DD2 model (thin) and FSU2H model (thick)  as a function of the density for $T=50$ MeV and $Y_Q=0.1$.  The $\Delta$ couplings are fixed to $ x_{\sigma \Delta}= x_{\omega \Delta}=1.2,~x_{\rho \Delta}=1$.	}
	\label{fig7}
\end{figure}

 In Fig.~\ref{fig6}, we plot together the fractions of unbound nucleons, light clusters, light hypernuclei, the $\Lambda$ fraction, the total $\Sigma$ fraction corresponding to the sum of the $\Sigma^{+,0,-}$ fractions, the total $\Xi$  fraction corresponding to the sum of the $\Xi^{0,-}$ fractions, the total $\Delta$ fraction corresponding to the sum of $\Delta^{-,0,+,++}$, for a charge fraction of $Y_Q=0.3$ (left) and 0.1 (right) in a calculation with (thick lines) and without (thin lines) $\Delta$s. The $\Delta$ couplings were chosen equal to the ones of the nucleons, $ x_{\sigma \Delta}= x_{\omega \Delta}=x_{\rho \Delta}=1$. We see that the inclusion of $\Delta$s  increases the abundances of the purely nucleonic light clusters above their maxima through the reduction of the binding energy shift of the clusters, so we would expect a similar increase for the hyperclusters. In fact, that is exactly what Fig.~\ref{fig6} shows. Once again, the effect is higher for the charge fraction $ Y_Q=0.1 $, since a smaller charge fraction favours negatively charged particles, which is the case of the $\Delta^-$ (the most abundant of the $\Delta$s). Therefore, if the $\Delta$s are more abundant for $ Y_Q=0.1 $, the reduction of the binding energy shifts of the hyperclusters after their maxima will be larger, resulting in higher dissolution densities and fractions. On the other hand, for densities below the hyperclusters maxima, the introduction of $\Delta$s actually slightly reduces the abundances of hyperclusters, which may  be due to a drop in the $\Lambda$s, which are essential to build hyperclusters.

\begin{figure*}%[tp]
	\includegraphics[width=0.9\linewidth]{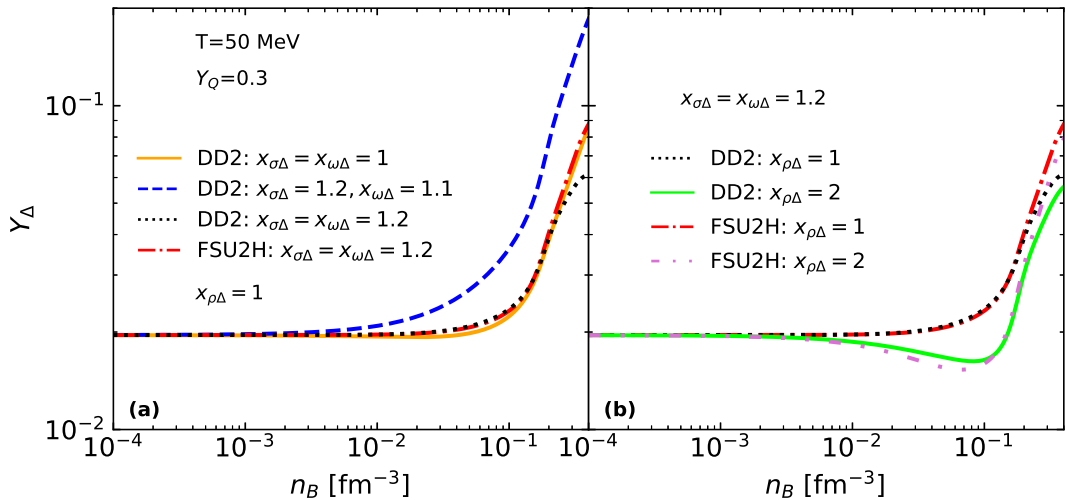}
	\caption{Total fraction of $\Delta$ isobars, $ Y_{\Delta} $ as a function of the density for $ T=50 $ MeV and a charge fraction of $ Y_{Q}=0.3 $. The $\sigma$-cluster meson fractions are $ x_s=0.93 $ (DD2) and $ x_s=0.91 $ (FSU2H), and several $\Delta$ couplings are considered. 	}
	\label{fig8}
\end{figure*}

In Fig.~\ref{fig7}, the particle fractions obtained with the models  DD2 (thin lines) and FSU2H (thick lines) are compared.  Although, the overall behavior is similar in both models, there are visible differences, in particular, after the peak of the cluster distributions: FSU2H model gives  smaller cluster fractions and  smaller dissolution densities. FSU2H also shows  smaller fractions of neutrons and hyperons but  larger $\Delta$ fractions. 
FSU2H favors the appearance of $\Delta$-baryons with respect to DD2, probably due to the difference on the $\rho$-meson couplings, see Eq.~\ref{Potential_Delta}. A higher $\Delta$ abundance for the FSU2H results in smaller fractions of hyperons compared to the DD2, especially negatively charged ones since the appearance of $\Delta^-$ disfavors negatively charged particles. The fractions of light clusters and hyper-clusters are also smaller for the FSU2H. This may be due to a smaller value of the fraction $x_s$ for the FSU2H ($x_s=0.91$) compared to the DD2 ($x_s=0.93$). In fact, the larger the $x_s$ the larger the $\sigma$-cluster couplings, resulting in a stronger binding, and therefore higher abundances.

Finally, let us now compare the total fraction of $\Delta$ isobars, $ Y_{\Delta} $, corresponding to the sum of $\Delta^{-,0,+,++}$ fractions, as a function of the density for a temperature $ T=50 $ MeV and a charge fraction of $ Y_{Q}=0.3 $, in a calculation  for different values of the $\Delta$ couplings using the DD2 and FSU2H models that includes unbound nucleons, hyperons, light clusters and hyperclusters.  In the left panel of Fig.~\ref{fig8}, we fix $ x_{\rho \Delta}=1 $ and perform the calculation for the three previously validated DD2 EoSs and the only valid FSU2H EoS. As we have mentioned before, the larger  $\Delta x=x_{\sigma \Delta} - x_{\omega \Delta}$, the higher the abundances of $\Delta$ isobars. On the other hand, DD2 parameterizations with  $\Delta x=0$ show similar abundances of $\Delta$s:  the  model with  higher $\sigma$ and $\omega$ couplings produces slightly higher abundances at smaller densities (where the $\sigma$ coupling is dominant \cite{Ribes_2019}) and lower abundances at higher densities (where the repulsion associated with the $\omega$ coupling dominates \cite{Ribes_2019}). As for the difference between the DD2 and FSU2H models with $ x_{\sigma \Delta}=x_{\omega \Delta}=1.2$, we can see that for small densities they show a similar fraction of $\Delta$s, whereas for higher densities the FSU2H starts yielding a higher fraction of $\Delta$s.
In the right panel of Fig. \ref{fig8}, we fix $ x_{\sigma \Delta}=x_{\omega \Delta}=1.2$ and perform the calculation for two values of $ x_{\rho \Delta}=1$ and 2 for DD2 and FSU2H. As we can see from Eq. (\ref{Potential_Delta}), the larger the value of $ x_{\rho \Delta}$, the less attractive the $\Delta^-$ potential is, making its presence less favorable, which is observed for both models. All these different parameterizations affect the fractions of the various particles, since a parameterization with more $\Delta$s than the one presented in Fig.~\ref{fig6} accentuates the effects mentioned in the discussion whereas a smaller abundance reduces the impact of the $\Delta$s.

The effect of heavy baryons on the presence of light clusters at low densities has also been discussed in Ref.\cite{Sedrakian2020}. In that study, the author includes pions, the $\Delta$-quadruplet and $\Lambda$ hyperons, besides nucleons and the classical light clusters ($ ^2 $H, $ ^3 $H, $ ^3 $He, $ ^4 $He). The
calculation is performed in the dilute limit within a Green's function formalism. Medium effects on the distribution of particles are included through the definition of the particle self-energies. For the nucleons, the self-energies are approximated by the nucleon effective masses and are calculated within a Skyrme nuclear matter model. A similar approach is introduced for the light clusters whose  self-energies are defined in terms of the nucleons effective masses. Besides, it is also included for clusters a temperature- and density-dependent  binding energy,   based on results from many-body calculations. The $\Lambda$ hyperon and the $\Delta$ isobars are taken with their vacuum masses and for the pions the leading contribution to the self-energy within a chiral
perturbation theory was considered. With the simplified description of the heavy baryons, the effect of the clusters on the heavy cluster fractions is not seen. In particular, the  heavy baryon fractions are insensitive to the cluster formations. Another effect is the fact that
in Ref. \cite{Sedrakian2020}, the $\Lambda$ fraction is larger  than the $\Delta$ fraction because they are defined by the baryon mass and the interaction with the medium is not considered. In our system, all particles interact with the medium in a self-consistent way, therefore the introduction of the heavy baryons such as hyperons and $\Delta$s does have an effect on the clusters abundance. Since our heavy baryons interact with the medium, their abundances do not depend only on their masses, which allows the $\Delta$ isobars to be more abundant than the $\Lambda$ hyperon for certain conditions.

In our study we did not include pions. Since the $\Delta$ isobars decay into a nucleon and pion through the strong force if there are available states, the presence of pions is expected in a finite temperature
scenario.  This will be analyzed in a future study.

\section{Conclusions \label{sec4}}

We have performed a calculation of clusterized matter including five light clusters ($ ^2 $H, $ ^3 $H, $ ^3 $He, $ ^4 $He,$ ^6 $He )
 and three light hyperclusters  ($ ^3_\Lambda $H, $ ^4_\Lambda $H, $ ^4_\Lambda $He), all hyperons belonging to the baryonic octet and the isospin multiplet of $\Delta$-baryons. The calculation was undertaken in the framework of relativistic mean-field theory, in particular, the models DD2 \cite{Typel2009} and FSU2H \cite{FSU2H} have been used.  Light clusters and hyperclusters were described as point-like particles, that interact with the mesons of the model, and besides feel a binding  energy shift due to the presence of the medium as introduced in \cite{Pais2018}. The binding energy shift is important to take into account in an effective way Pauli blocking effects. The hyperon-meson couplings were obtained from a calibration to hypernuclear properties \cite{Fortin_Hypernuclei_and_massive_neutron_stars}. 
  In order to choose adequate $\Delta$-meson couplings we have  considered experimental constraints  as summarized in \cite{Drago2014} and imposed  as well observational constraints. 
 Many parametrizations  for the $\Delta$-baryons had to be disregarded because nucleon effective masses would become zero before a maximum star mass would be reached. 
 
 The main conclusions of the present work are: (i) the presence of heavy-baryons, both hyperons and $\Delta$s favour the formation of clusters and shift their dissolution to larger densities; (ii) a larger number of clusters decreases the fraction of free nucleons, and, in particular, the difference between the fractions of neutrons and protons decrease, which favors processes like direct Urca reactions. In the future, it is important to implement the presence of heavy baryons in the low density warm  EoS used in core-collapse supernova or neutron star merging simulations.

\section*{ACKNOWLEDGMENTS}
This work was partly supported by the FCT (Portugal) Projects No. UIDB/FIS/04564/2020, UIDP/FIS/04564/2020 and POCI-01-0145-FEDER-029912, and by PHAROS COST Action CA16214. H.P. acknowledges the grant CEECIND/03092/2017 (FCT, Portugal).

\end{document}